\documentclass[pra,groupedaddress,superscriptaddress,twocolumn,showpacs,amsmath,amssymb,a4paper]{revtex4}
\usepackage{geometry}
\usepackage{graphicx}
\usepackage{hyperref}
\usepackage{dsfont}
\usepackage{amssymb}
\usepackage{amsmath}
\usepackage{setspace}
\usepackage[T2A]{fontenc}
\usepackage[utf8x]{inputenc}
\usepackage[russian,english]{babel}
\usepackage{braket}
\graphicspath{{pic/}}
\begin{document}
\title{Mean propagation velocity of multiphoton wave-packet states with nonzero Lorentz-invariant mass.}

\author{S. V. Vintskevich }

\affiliation{Moscow Institute of Physics and Technology,
Institutskii Per. 9, Dolgoprudny, Moscow Region 141700, Russia}
\author{D.A. Grigoriev}

\affiliation{Moscow Institute of Physics and Technology,
Institutskii Per. 9, Dolgoprudny, Moscow Region 141700, Russia}
\begin{abstract}
The concept of Lorenz invariant mass and mean propagation velocity have been investigated in detail for various multiphoton wave-packet states of light. Based on photodetection theory and straightforward kinematics, we presented a physically reasonable and at the same time rigorous proof that mean propagation velocity is consistent with the Lorentz-invariant mass concept for an arbitrary multiphoton wave-packet state. We argued that mean propagation velocity is less than the speed of light constant in vacuum and is governed by geometric properties of state's amplitude in wave-vector space for arbitrary wave-packet states. To classify states with different fixed values of Lorentz-invariant mass, we introduced a specific set of modes that allow us to describe the wave-packet state in its rest frame.
\end{abstract}


\maketitle

\section{Introduction}
Present-day experimental techniques in quantum optics allow measuring the time of arrival (ToA) of single photons with femtosecond precision. Quantum interference phenomena play an essential role in such measurements. In particular, the Hong-Ou-Mandel (HOM) effect \cite{HOM} was utilized to detect the speed of propagation of structured light photons in free space, which was shown to be less than the speed of light \cite{giovannini}. We believe this result to have an important methodological meaning.  It follows that practical implementation of "exotic" (structured) photon states requires a reexamination of some of properties of such states related to propagation speed, even in vacuum \cite{Saari2018, Saari2020,Lyons,Bouchards,Petrov2019}.

Recently, a new approach has been proposed, which is based on a somewhat unusual implementation of the well known physical concept of Lorentz-invariant mass (LI mass). The concept of LI-mass is directly related to mean propagation velocity of  classical light pulses, including structured, \cite{LP2017,vints2019,WE_2017_EPL} and multiphoton quantum states \cite{LPL2019}. 

The physical meaning of light pulse's or state's LI mass can be described by the following statements:

1. If a system has nonzero LI mass, there exists a reference frame where the system as a whole remains at rest. For a spatially localized state of light, it means that spreading is the only motion existing in this reference frame.

2. LI mass of a light state (light pulse) characterizes the energy of its diffraction.

3. Let $\braket{H}$ and $\braket{\vec{p}}$ be mean energy and momentum of light in the laboratory reference frame. One can calculate LI mass and mean propagation velocity as follows:
\begin{eqnarray}\label{speed}
&& m^2c^4 = \braket{H}^2 - \braket{\vec{p}}^2c^2
\nonumber\\
&& \frac{v}{c} = \sqrt{1-\frac{m^2c^4}{ \braket{H}^2}} = \frac{c|\braket{\vec{p}}|}{\braket{H}}
\end{eqnarray}
In our previous work \cite{vints2019}, we showed that any structured localized light pulses propagate with the speed that is less than the speed of light in vacuum. Moreover, the mean propagation velocity does not depend on the pulse energy, but on its spatial structure, i.e., its distribution in the wave-vector space. This claim is a direct inference of the third statement and \eqref{speed} above. On the other hand, according to the first statement, there exists a reference frame where a given structured pulse, and even more generally, any spatially localized state of light, is at rest as a whole. This observation leads to natural questions:  can one provide a rigorous derivation of the slowing down effect for arbitrary states based on photons registration (light intensity) in the experiment?
How can we classify the state of photons with a fixed value of LI mass? This paper aims to give a supplement to listed LI mass fundamental physical meaning, answering the above questions. We provide several illustrative examples in Section \ref{sec2} to give proper intuition and perspective on LI mass of various quantum states. The presented derivation does not imply any {\it{a priori}} information about the reference frame. Contrary, in Section \ref{sec4}, we focus on a rest frame specific property and demonstrate how to use spherical harmonic modes to represent multiphoton states with nonzero LI mass. Modes with well-defined LI mass are constructed based on the wave-packet quantization technique proposed by U. M.  Titulaer and R.J. Glauber \cite{TG} and further developed by B.J. Smith and M.G. Raymer \cite{Smith} within a problem of a photon's wave function in coordinate representation (we also recommend \cite{IWO,KELLER}). In conclusion, we briefly discuss specific methods and tools which we believe could be useful in possible experiments to detect various effects, in particular, related to LI mass.
\section{Calculation of the Lorentz-invariant mass for various states}\label{sec2}
One can calculate LI mass for an arbitrary quantum state of light described by density operator $\hat{\varrho}$ as follows:
\begin{equation}\label{mass_general}
m^2c^4 = \left({\rm{tr}}\hat{\varrho} \hat{H}\right)^2 - \left({\rm{tr}}\hat{\varrho} \hat{\vec{p}}\right)^2c^2,
\end{equation}
where ${\rm{tr}}$ is trace operation. One Density operator $\rho$ can be represented either in the Fock basis or in the basis of multimode coherent states \cite{Mandel_Wolf}; $\hat{H} = \int d\vec{k}\hbar \omega_{k}\hat{a}^{\dagger}_{s}(\vec{k})\hat{a}_{s}(\vec{k})$ and $\hat{\vec{p}} = \int d\vec{k}\hbar \vec{k}\hat{a}(\vec{k})^{\dagger}\hat{a}(\vec{k})$ are Hamiltonian and momentum operators of free electromagnetic field. Operators  $\hat{a}_{s}(\vec{k})$ and  $\hat{a}^{\dagger}_{s}(\vec{k})$ are annihilation and creation operators. The subscript $s$ denotes polarization degree of freedom,
$\omega_{k} = c\sqrt{k_{x}^2+k_{y}^2+k_{z}^2}$ is the dispersion relation for electromagnetic field in vacuum in accordance with Maxwell equations, $c$ is the constant of light speed in vacuum. 
\subsection{LI mass of two mode pure and mixed states.}
To make the derivation more clear, we suppose that the polarization is fixed, focusing our attention on plane-wave modes of photons with a well-defined wave-vector $\vec{k}$.
We will begin with basic example of single-photon pure Fock state $\ket{1_{\vec{k}_0}}$.
LI mass equals zero for this state as expected according to \eqref{mass_general}. Moreover, LI mass is also zero in the case of $n$-photon Fock state with fixed mode $\vec{k}_0$. In contrast, one can consider another $n$-photon state:
\begin{equation}\label{simpliest}
\ket{\psi} = \ket{\frac{n_{\vec{k}_{1}}}{2}, \frac{n_{\vec{k}_{2}}}{2}},\ n_{\vec{k}_{1}} = n_{\vec{k}_{2}} = n \geq 2,
\end{equation}
which is a Fock state with two different $k$-modes ($\vec{k}_1$ and $\vec{k}_2$) and $n/2$ occupation number in each mode. $n$ is assumed to be even positive integer. For the sake of simplicity we choose $|\vec{k}_1| = |\vec{k}_2| = \omega_{0}/c$. The state \eqref{simpliest} has nonzero LI mass:
\begin{equation}\label{n-ph-lim}
m_{n-\rm{ph}} = \frac{n\hbar\omega_{0}}{c^2}\sin{\vartheta/2},
\end{equation}
where $\vartheta \in \left(0,\pi \right]$ is the angle between wave vectors $\vec{k}_1$ and $\vec{k}_2$. This equation is the direct $n$-photon generalization of the expression, obtained in the work \cite{Okun} by L.B. Okun for two-mode states in case of $n=2$ (two noncollinear photons) and also mentioned by P.Saari in \cite{saari_book}. Note that if $\vartheta =0$ in \eqref{n-ph-lim} which means collinear propagation of $n$-photons, LI mass is zero as expected. 

Let us now consider LI mass of an another important model of mixed quantum  $n$-photon states and compare it with its pure state. We choose the following density matrix and $n$-photon state to describe this $n$-photon mixed state:
\begin{eqnarray}\label{mixed_state}
&&\hat{\varrho}_{n-{\rm{ph}}} = \sum_{i} \lambda_{i}\ket{n_{\vec{k}_{i}}}\bra{n_{\vec{k}_{i}}}, \  \sum_{i} \lambda_{i} = 1 \\
&& \ket{\psi}_{n-{\rm{ph}}} = \sum_{i}\sqrt{\lambda_{i}}e^{i\phi_{i}}\ket{n_{\vec{k}_{i}}},
\label{mixed_state_1}
\end{eqnarray}
where $\phi_{i}$ is a phase of state with wave vector $\vec{k}_{i}$ into the coherent superposition. 
Again, we assume $|\vec{k}_i| = \omega_{0}/c, \ \forall \ i $ similar to the above. Let us denote $\vartheta_{ij}$ to be the angle between wave vectors $\vec{k}_{i}$ and $\vec{k}_{j}$, and  $\vartheta_{ij} = 0$ if $i=j$.  The mean energy in case of \eqref{mixed_state} and \eqref{mixed_state_1} according to \eqref{mass_general} reads:
\begin{eqnarray}\label{mass_mixed}
{\rm{tr}}(\hat{\varrho}_{n-{\rm{ph}}}\hat{H}) = \sum_{i} \lambda_{i} \int d\vec{k}\hbar \omega_{k}\bra{n_{\vec{k}_{i}}}\hat{a}_{\vec{k}}^{\dagger}\hat{a}_{\vec{k}}\ket{n_{\vec{k}_{i}}} = n\hbar\omega_{0}. \\
{\rm{tr}}(\ket{\psi_{n-{\rm{ph}}}}\bra{\psi_{n-{\rm{ph}}}}\hat{H}) = \nonumber \\
\int d\vec{k}\hbar \omega_{k}\sum_{i,j}\sqrt{\lambda_{i}\lambda_{j}}e^{i(\phi_{j}-\phi_{i})}\bra{n_{\vec{k}_{i}}}\hat{a}_{\vec{k}}^{\dagger}\hat{a}_{\vec{k}}\ket{n_{\vec{k}_{j}}}= \nonumber\\
 =n\hbar\omega_{0}
\label{mass_mixed_1}
\end{eqnarray}
The momentum calculation is also straightforward:
\begin{eqnarray}
{\rm{tr}}(\hat{\varrho}_{_{n-{\rm{ph}}}}\hat{\vec{p}}) = {\rm{tr}}(\ket{\psi_{n-{\rm{ph}}}}\bra{\psi_{n-{\rm{ph}}}}\hat{\vec{p}}) = n\hbar\sum \lambda_{i}\vec{k_{i}}
\end{eqnarray}
As a result, according to \eqref{mass_general} calculation of LI mass yields:
\begin{eqnarray}\label{mass_mixed2}
&&m^2c^4 = (n\hbar\omega_{0})^2\left(1 - \sum_{i,j}\lambda_{i}\lambda_{j}\cos\left(\vartheta_{ij}\right)\right) \implies \nonumber \\
&&m = \frac{2n\hbar \omega_{0}}{c^2}\sqrt{\sum_{i>j}\lambda_{i}\lambda_{j}\sin^2\left(\frac{\vartheta_{ij}}{2}\right)}.
\end{eqnarray}
In general, LI mass of $n$-photon mixed and pure states presented in \eqref{mass_mixed} and \eqref{mass_mixed_1} does not coincide with pure two mode $n$-photon state in \eqref{simpliest}. 
But, formally, if $\lambda = 1/2$, LI mass is the same in both \eqref{n-ph-lim} and \eqref{mass_mixed2}.
It is clear from examples above that coherent and incoherent superposition gives the same mean value of LI mass, but of course, one can easily distinguish these states e.g., by performing HOM based measurements \cite{Cassemiro2010}.

However, the case of a mixed $n$-photon state is not that simple. The density matrix in \eqref{mixed_state} defines the statistical properties of an ensemble, where coefficients $\lambda_{i}$ are {\it{classical}} probabilities. This means that we prepare the $n$-photon state with a specific wave-vector from one experimental realization to another. For any particular state presented in the mixture \eqref{mixed_state}, LI mass is zero. In contrast, the coherent superposition (the pure state in \eqref{mixed_state_1} ) describes a multi-photon wave packet, which can be treated as an isolated system. In other words, in every experimental realization, we prepare the same pure state, which has nonzero LI mass. Thus, a state (ensemble) preparation procedure plays a crucial role in physical interpretation of LI mass of a quantum state of light. But, one may expect that lack of difference between LI mass of different mixed and pure states is inevitable, because LI mass was defined through already averaged Hamiltonian and momentum operators. 

In modern experiments, one can prepare single-photon temporal wave-packet states by conditional measurement using spontaneous parametric down-conversion (SPDC) two-photon source. One can also produce single-photon wave-packets in experiments with spontaneous emission of an atom (we recommend, e.g. \cite{Lounis2005}). It is worth to mention that SPDC is well-known and standard tool to prepare two-photon entangled wave-packets. Nevertheless, we should keep in mind that the effects related to LI mass stem from spatial shaping and localization of multiphoton wave-packet states, which we will consider in the next section.

\subsection{LI mass of wave-packet states.}
We pointed out in the previous section that LI mass of a single or $n$-photons Fock state describing state with the fixed mode $\vec{k}$ is zero. But in fact, it is not the case for single photon wave-packet. 
Let us consider the following state:
\begin{eqnarray}\label{wp_1}
&&\ket{\psi}_{{\rm{wp}}} = \sum_{i}\psi_{i}\ket{1_{\vec{k}_{i}}} = \sum_{i}\psi_{i} \frac{\hat{a}_{\vec{k}_i}^{\dagger}}{\Delta k^{\frac{3}{2}}}\Delta k^{\frac{3}{2}}_{i}
\implies \nonumber \\
&& \int \psi(\vec{k})\hat{a}^{\dagger}(\vec{k})\ket{vac}d\vec{k} =  \int \psi(\vec{k})\ket{1_{\vec{k}}}d\vec{k}; \  \nonumber \\
&& \ \ \ \ \ \ \ \ \ \ \ \ \ \  \hat{a}^{\dagger}(\vec{k}) = \frac{\hat{a}_{\vec{k}_i}}{\Delta k^{\frac{3}{2}}}, \ \Delta k_{i}\longrightarrow 0,
\end{eqnarray}
where $\ket{1_{\vec{k}_{i}}}\equiv \ket{\dots 000 \vec{k}_{i} 000 \dots}$ is a single mode Fock state with only $\vec{k}_{i}$ mode exited, $\Delta{k}^{3/2}_{i} = (2\pi/L)^{3/2}$, $\hat{a}_{\vec{k}_i}^{\dagger}$ and $\hat{a}^{\dagger}(\vec{k})$ are the photon creation operators in discrete and continuous space representation respectively. Here we use a standard transition from discrete summation to integral representation with continuous variable $\vec{k}$:
$(L/2\pi)^{3}\sum_{k_{i}}\Delta_{{k}_{i}}\longrightarrow \int d\vec{k}$, assuming that volume element $\Delta k ^3$ in $\vec{k}$ - space tends to zero. Also we replaced $\psi_{i}$ with its continuous variable analog $\psi\left(\vec{k}\right)$.
We can easily generalize state in \eqref{wp_1} and produce $n$-photon wave-packet state probabilistically using for e.g. beamsplitters (comprehensive review and methodological insight are given by Z.Y.J.Ou \cite{ou2007multi,ou2017quantum}):
\begin{equation}
    \ket{\psi}^{{\rm{n-ph}}}_{\rm{wp}} = \int \Psi(\vec{k})\ket{n_{\vec{k}}}d\vec{k}
\end{equation}
Let us calculate LI mass for one specific state \cite{Mandel_Wolf}, where state $\ket{\phi}^{{\rm{n-ph}}}_{\rm{wp}}$ corresponds to the pseudo localized single photon state, center at given time in point $\vec{r}_{0}$: \begin{equation}\label{wp-mw}
\ket{\phi}^{{\rm{n-ph}}}_{\rm{wp}} = \int \frac{1}{(\sqrt{\pi}\sigma)^{\frac{3}{2}}} e^{-\frac{\left(\vec{k}-\vec{k}_{0}\right)^2}{2\sigma^2}e^{i\vec{k}\vec{r_{0}}}} \ket{1_{\vec{k}}}d \vec{k}.
\end{equation}
One can immediately check that average number of photons in the state \eqref{wp-mw} is $\braket{n} = \int \braket{\phi \hat{a}^{\dagger}(\vec{k})\hat{a}(\vec{k})\phi}d\vec{k} \equiv 1 $ , and state is properly normalized. Let us substitute \eqref{wp-mw} in \eqref{mass_general}. At first we calculate the mean momentum, choosing  $z$ - coordinate axis along  $\vec{k}_{0}$. To calculate mean energy we use spherical coordinates in $\vec{k}$ - space ($k, \theta, \phi$). As a result we have:
\begin{eqnarray}
\label{wp_momentum}
&&\braket{\hat{\vec{p}}} = \hbar \vec{k}_{0} \nonumber \\ 
&&\braket{\hat{H}} = \frac{\hbar c}{(\sqrt{\pi} \sigma )^3} \int k^3e^{\frac{-\left(\vec{k}-\vec{k}_{0}\right)^2}{\sigma^2}}\sin{\theta} dkd\theta d\phi = \nonumber \\
&& \frac{2\pi \hbar c e^{-\frac{\vec{k}_{0}^2}{\sigma^2}}}{(\sqrt{\pi} \sigma )^3}\int_{0}^{\infty} k^3e^{-\frac{\vec{k}^2}{\sigma^2}}\int_{0}^{\pi}e^{\frac{2kk_{0}\cos{\theta}}{\sigma^2}}\sin{\theta}d\theta dk = \nonumber \\
&& = \hbar c k_{0}{\rm{erf}}\left(\frac{k_{0}}{\sigma}\right)\left(1 +\frac{\sigma^2}{2k_{0}^2}\right)+\sqrt{\frac{1}{\pi}} \hbar c \sigma e^{-\frac{k_{0}^2}{\sigma^2}}
\end{eqnarray}
Finally, using calculated mean momentum and energy  \eqref{wp_momentum} we get "not obvious" result that LI mass is not zero for the single-photon wave packet:
\begin{eqnarray}\label{masswp}
&&m^2c^4 =\left(\hbar c k_{0}{\rm{erf}}\left(\frac{k_{0}}{\sigma}\right)\left(1 +\frac{\sigma^2}{2k_{0}^2}\right)+\sqrt{\frac{1}{\pi}} \hbar c \sigma e^{-\frac{k_{0}^2}{\sigma^2}}\right)^2 \nonumber \\
&&-\left(\hbar c k_{0}\right)^2 \approx \left(hc\sigma\right)^2, k_{0}\gg \sigma \implies \nonumber \\
&& m \approx \frac{\hbar \sigma}{c}.
\end{eqnarray}
The obtained result is definitely in agreement with the second statement  in the introduction regarding LI mass's physical sense. LI mass is nonzero for wave-packets with finite spatial size. Here we provide an example from paper \cite{LPL2019} which supports this observation. In this paper, the authors considered two-photon (another name is biphoton) wave packet produced in a frequency - degenerate Type-I SPDC process. This two-photon state is entangled and has the following form: 
\begin{eqnarray}\label{biph}
\ket{\phi}_{\rm{biph}} \propto \int d\vec{k}_{{\perp 1}}d\vec{k}_{{\perp 2}} e^{-\left(\vec{k}_{\perp 1}+\vec{k}_{\perp 2}\right)^2 w_{p}^2}\times\nonumber\\
\times{\rm{sinc}}{\left[\frac{L\lambda_{p}}{8\pi n_{o} }\left(-(\vec{k}_{\perp 1}+\vec{k}_{\perp 2})\right)^2\right]}\ket{1_{\vec{k}_{{\perp 1}}}, 1_{\vec{k}_{{\perp 2}}}},
\end{eqnarray}
where L is the length of a crystal (along the $z$-axis), $n_{o}$ is the refractive index of the ordinary wave in the crystal, $w_{p}$ and $\lambda_{p}$ are waist and central wavelength of a pump beam. Moreover, authors established direct link between entanglement measure and LI mass of biphoton state \eqref{biph}. It was shown, that if $w_{p}\gg \sqrt{L\lambda_{p}}$, the Lorentz-invariant mass of biphoton pairs can be rewritten as
\begin{equation}\label{mass-via-K}
    m_{\rm biph}\approx\frac{\hbar K}{2 c\,w_p}\sqrt{\frac{1}{\pi}\ln\left(\frac{\pi L}{2n_o\lambda_p}\right)}\gg\frac{\hbar}{2 c\,w_p},
\end{equation}
where constant $K \propto \frac{2\pi w_{p}\sqrt{n_{0}}}{\sqrt{L\lambda_{p}}} \gg 1$ is Schmidt number characterizing the degree of entanglement of the biphoton state. One can see that again, wave-packet's spatial size is the key parameter that governs LI mass along with spatial mode structure. Thus, in general, LI mass of an arbitrary wave-packet state can be roughly estimated as follows:
\begin{equation}\label{mass_evr}
m \propto \frac{\hbar}{c}\frac{\rm{const}}{\sqrt[3]{V}},
\end{equation}
where $const$ depends on wave-packet structure defined by state's amplitude $\psi(\vec{k}, \dots)$, $V$ is a localization volume in the coordinate space. This result agrees with the purely classical approach for spatially localized light pulses \cite{vints2019,WE_2017_EPL}. To conclude, it is worth to calculate the mean propagation velocity for the wave-packet  \eqref{wp-mw}. According to definition \eqref{speed} mean propagation velocity of \eqref{wp-mw} is given by:
\begin{eqnarray}
v = c\sqrt{\left(1 - \frac{\sigma^2}{k_{0}^2}\right)},
\end{eqnarray}
which is in accordance with experimental results in \cite{giovannini} and classical derivations in \cite{WE_2017_EPL,LP2017,vints2019}. But, we would like to put attention on fact, that overall result does not depend on Plank constant $\hbar$ which supports the idea that propagation velocity is defined by geometrical properties of a mode, but LI mass represent joint corpuscular-wave (excitation-mode) nature of a spatially localized photon state. 
\subsection{Remark about a polarization degree of freedom and  calculation of wave-packet's LI mass.}
In the examples illustrated above, we excluded polarization degrees of freedom from consideration, which, of course, gives an incomplete scope of LI mass's properties. In this subsection, we consider wave-packet states, including the polarization degree of freedom.
Let us consider $n$-photon wave-packet state:
\begin{eqnarray}\label{WP_POL}
\ket{\psi}^{\rm{pol}}_{\rm{wp}} = \sum_{s}\int \psi_{s}\left(\vec{k}\right)\ket{n_{\vec{k},s}} d\vec{k},
\end{eqnarray}
where index $s$ stands for the polarization degree of freedom, and we assume the following normalization condition for the wave function $\sum_{s}\int d \vec{k} |\psi_{s}\left(\vec{k}\right)|^2 = 1$. To calculate mean energy, momentum and LI mass according to \eqref{mass_general} we denote $|\psi(\vec{k})|^2$ as an analog of marginalized probability $|\psi(\vec{k})|^2 = (1/\braket{n})\sum_{s} n_{s} |\psi_{s}\left(\vec{k}\right)|^2$  with respect to the polarization variable $s$, but normalized by average number of photons, where $\braket{n} = \int \sum_{s} n_{s}|\psi_{s}\left(\vec{k}\right)|^2 d\vec{k}$. As a result we have:
\begin{eqnarray}\label{mom-en-polwp}
&& \braket{\hat{H}} =\sum_{s}\int \hbar \omega_{\vec{k}}\braket{\psi^{\rm{pol}}_{\rm{wp}}a_{s}^{\dagger}(\vec{k})a_{s}(\vec{k})\psi^{\rm{pol}}_{\rm{wp}}}d\vec{k} = \nonumber \\
&& = \braket{n}\int \hbar \omega_{\vec{k}} |\psi\left(\vec{k}\right)|^2 d \vec{k} \nonumber \\
&&  \braket{\hat{\vec{p}}} =   \braket{n}\int \hbar \vec{k}|\psi\left(\vec{k}\right)|^2 d \vec{k},
\end{eqnarray}
The LI mass is calculated as usual, based on \eqref{speed}.
However, if one considers a coherent superposition of two wave-packet states with different amplitudes which, in general, depend on the polarization index: $\tilde{\phi}_{s}(\vec{k})$ and $\phi_{s}(\vec{k})$, consequently resulting LI mass must be expressed through $\sum_{s}|\tilde{\phi}_{s}(\vec{k}) \pm \phi_{s}(\vec{k})|^2$.  Such sum of amplitudes gives rise of possible interference effects which influence on LI mass and mean propagation speed as well. We believe this straightforward observation may be useful for a possible experiments involving LI mass.
\section{Derivation of n-photon wave-packet's mean propagation velocity based on intensity photodetection} \label{sec3}
We learned about LI mass's properties in the previous section, but we did not provide any information about how to detect the effects related to LI mass, namely the slowing down effect. In this section, we are answering the first question from the introduction, regarding mean propagation velocity. Namely, can one provide a rigorous derivation of the slowing down effect for arbitrary states based on photons registration (light intensity) in the experiment? From \eqref{speed} and \eqref{mass_general} one can easily note that calculations of mass and mean propagation velocity are directly related to finding of quantum mechanical averages of operators, which depend on linear combinations (with respect to mode index $\vec{k}$) of normally ordered operators $\hat{a}^{\dagger}_{s}\left(\vec{k}\right)\hat{a}_{s}\left(\vec{k}\right)$, which in turn are directly related to the field intensity in the experiment. On the other hand, in experiment one observes detector's rates (counts) of the registering photon. According to the R.J. Glauber's photodetection theory \cite{Glauber1}, to describe intensity measurements for one-photon state one needs to calculate the first order field correlation function:
\begin{eqnarray}
&&\Gamma(\vec{r}_{1},t_{1},\vec{r}_{2},t_{2}) = \braket{\Psi\hat{\vec{E}}^{(-)}(\vec{r}_{1},t_{1})\hat{\vec{E}}^{(+)}(\vec{r}_{2},t_{2})\Psi},
\end{eqnarray}
where $\hat{\vec{E}}^{+}(\vec{r},t_{1})$ is the positive frequency electric field operator. Here we use the standard flat-wave mode expansion of field operators:
\begin{eqnarray}\label{operator}
\hat{\vec{E}}^{+}(\vec{r},t) = i\frac{\sqrt{\hbar c}}{2\pi}
\sum_{s}\int d\vec{k}\hat{a}_{s}(\vec{k})\sqrt{k}
\vec{\epsilon}_{\vec{k},s}e^{i\vec{k}\vec{r} - \omega_{k}t}.
\end{eqnarray}
Note that $\hat{\vec{E}}^{(-)}(\vec{r},t) = \hat{\vec{E}}^{{(+)}{ \dagger}}(\vec{r},t)$. Let us consider the case of an arbitrary $n$-photon wave-packet state introduced in \eqref{WP_POL}. One can generalize this result further by considering a suppositions of wave-packet states given by \eqref{WP_POL} with different number of photons and various forms of amplitudes (wave-functions) $\psi_{s}\left(\vec{k}\right)$. It is worth to mention that one can construct an even more general state by including the superposition of $n$-photon states.
Assume that we place an array of ideal (with sizes much less then wave-packet's characteristic size) detectors in an arbitrary plane $z$, where we suppose that $z\gg c\overline{\tau}$ and $\overline{\tau}$ is wave-packet temporal width. Let us choose an arbitrary point $\vec{r} = \vec{r}_{\perp} + z\vec{e}_{z}$. Thus registered intensity distribution at this point and time interval $t+dt$ reads: 
\begin{eqnarray}\label{intensity distr}
p_{z}\left(\vec{r}_{\perp},t\right) = \frac{\Gamma(\vec{r},t,\vec{r},t)}{\int  \Gamma(\vec{r},t,\vec{r},t) d\vec{r}_{\perp}dt},
\end{eqnarray}
and
\begin{eqnarray}\label{velocity_int}
\braket{v} =|\braket{\frac{\partial \vec{r}}{\partial t}}| = |\int \frac{\partial \vec{r}}{\partial t}p_{z}\left(\vec{r}_{\perp},t\right)d\vec{r}_{{\perp}}dt|
\end{eqnarray}
The detailed derivation and proposed simplifications are given in Appendix (\ref{appendix}). In this section we leave only key expressions and calculation results.
Firstly, one can easily find that denominator in \eqref{intensity distr} is equal $2\pi\braket{\hat{H}}$ which in fact is the energy normalization of intensity distribution.
Secondly, substituting in \eqref{velocity_int} expressions \eqref{intensity distr}, \eqref{operator} we have to calculate the following expression:
\begin{eqnarray}\label{derivatives-sec3}
&&\bra{\Psi}\int \frac{\partial \vec{r}}{\partial t}\hat{\vec{E}}^{(-)}(\vec{r},t)\hat{\vec{E}}^{(+)}(\vec{r},t) d\vec{r}dt\ket{\Psi} = \nonumber \\
&&\bra{\Psi}\int \frac{\partial}{\partial t}\left(\vec{r}\hat{\vec{E}}^{(-)}(\vec{r},t)\hat{\vec{E}}^{(+)}(\vec{r},t)d\vec{r}\right)dt - \nonumber \\
&& \int \vec{r}\left(\frac{\partial  \hat{\vec{E}}^{(-)}(\vec{r},t)\hat{\vec{E}}^{(+)}(\vec{r},t)}{\partial t}dt\right) d\vec{r}\ket{\Psi}
\end{eqnarray}
Long but straightforward calculations yield that the overall expression reduces to the average momentum multiplied by the speed of light constant $c$: 
\begin{eqnarray}
&&\int \frac{\partial}{\partial t}\left(\vec{r}\braket{\Psi\hat{\vec{E}}^{(-)}(\vec{r},t)\hat{\vec{E}}^{(+)}(\vec{r},t)\Psi}d\vec{r}\right)dt = \nonumber\\
&&c\int (k_{\perp}\vec{e}_{\perp}+k_{z}\vec{e}_{z}) \sum_{s} |\psi_{s}\left(\vec{k}\right)|^2 d\vec{k} = c\braket{\hat{\vec{p}}} 
\end{eqnarray}
Combining all the calculation results, we proved the validity of the general formula for mean propagation velocity \eqref{velocity_int} and obtain the following final expression:
\begin{eqnarray}
\braket{v} =|\braket{\frac{\partial \vec{r}}{\partial t}}| = c\frac{|\braket{\hat{\vec{p}}}|}{\braket{H}}.
\end{eqnarray}
This derivation demonstrates that the straightforward kinematic approach is in agreement with the dynamic treatment and concept of LI mass, answering the first question in the introduction. 
However, it is important to emphasize that we made averaging over intensity distribution recorded by array of detectors (pixels) of the ideal bucket detector absorbing the whole $n$-photon state. Moreover, one can perform somewhat different analysis based on higher orders of correlation functions and different setups, which could lead to other fruitful results.

\section{Classification of states with fixed Lorentz-invariant mass} \label{sec4}
General formulas of LI mass that we gave in previous sections for various states make sense only if $\braket{\Delta k_{x}\Delta k_{y}\Delta k_{z}}\neq 0$, or in other words, if the state has nonzero volume in $k$-space and in the coordinate space consequently. If this requirement is met, we have a wave packet with the real-valued nonzero LI mass. However, it is clear that in general, different multiphoton states could have the same value of LI mass. In this section we try to construct sets of states with fixed LI mass. According to the physical meaning of LI mass, it is the energy of the wave-packet (localized light pulse) in its rest frame. Consequently, the most direct way to classify states with fixed LI mass is to describe an arbitrary wave-packet in its reference frame. Thus, it means that we impose the following restrictions:

{\it{(i)}} The mean value of momentum operator for any of these special modes is zero: $\braket{\hat{\vec{p}}} \equiv 0$. 
Mathematically we impose that:
\begin{eqnarray}\label{mom-en-polwp-null}
&& \braket{\hat{H}} =\sum_{s}\int \hbar \omega_{\vec{k}}\braket{{\psi}^{\rm{r.f}}_{\rm{wp}}a_{s}^{\dagger}(\vec{k})a_{s}(\vec{k}){\psi}^{\rm{r.f}}_{\rm{wp}}}d\vec{k} = \nonumber \\
&& = \int \hbar \omega_{\vec{k}}\left(\sum_{s}  |\psi_{s}^{\rm{r.f}}\left(\vec{k}\right)|^2\right) d \vec{k} = mc^2 \nonumber \\
&&  \braket{\hat{\vec{p}}} = \int \hbar \vec{k}\left(\sum_{s}  |\psi_{s}^{\rm{r.f}}\left(\vec{k}\right)|^2\right) d \vec{k} = 0
\end{eqnarray}
where $r.f.$ - indicates that we consider some specific set of states $\ket{\{\psi^{r.f.}\}}$ in the rest frame.

{\it{(ii)}} Each mode represents wave-packet state with nonzero LI mass in its rest frame. 

{\it{(iii)}} We assume that the field is free and influence of the source is insignificant.

Let us now construct such modes that would satisfy the above conditions. The requirement $\braket{\hat{\vec{p}}} \equiv 0$ imposes that these new modes must express rotational symmetry properties. The most natural candidates seem to be the spherical harmonics \cite{cohen2006quantum}, which are the basis functions for irreducible representations of SO(3) rotational group in three dimensions
\begin{eqnarray}
&&Y_{l,j}\left(\theta_{\vec{k}},\phi_{\vec{k}}\right) = (-1)^{j}\sqrt\frac{(2l+1)(l-j)!}{4\pi (l+j)!}P_{l}^{j}({\rm{cos}}(\theta))e^{i(j\phi_{\vec{k}})},\nonumber \\
&&j\geq 0.
\end{eqnarray}
where $P_{l}^{j}({\rm{cos}}(\theta))$ - are associated Legendre polynomials.
These functions form a complete set of orthonormal functions :
\begin{eqnarray}\label{completeort}
&&\sum_{l,j}Y_{l,j}\left(\theta_{\vec{k}},\phi_{\vec{k}}\right)Y_{l,j}^{*}\left({\theta'}_{\vec{k}},{\phi'}_{\vec{k}}\right) = \frac{\delta(\theta_{\vec{k}} - {\theta'}_{{k}})\delta(\phi_{\vec{k}} - {\phi'}_{\vec{k}})}{\sin{(\theta_{\vec{k}})}}; \nonumber \\
&& \int_{0}^{2\pi}\int_{0}^{\pi} d\phi_{\vec{k}}d\theta_{\vec{k}}\sin(\theta_{\vec{k}})Y_{l,j}\left(\theta_{\vec{k}},\phi_{\vec{k}}\right)Y_{l,j}^{*}\left({\theta}_{\vec{k}},{\phi}_{\vec{k}}\right)=\nonumber \\
&&= \delta_{ll'}\delta_{jj'}
\end{eqnarray}
As we mentioned earlier, LI mass is the energy of spreading in the rest frame, thus one needs to expect that its value must be related to Hamiltonian's eigenvalues - energy. In other words, states with fixed LI mass are degenerate. This observation concurs with rotational symmetry and one imposed by requirement $ \braket{\hat{\vec{p}}} \equiv 0$.
To construct this set of states we use the approach proposed by Titulaer and Glauber \cite{TG} and more recently extended by Smith and Raymer \cite{Smith} by introducing 
new set of modes. Thus, let us define new annihilation and creation operators for a single photon, or one may say  - excitation quanta, in this new modes: 
\begin{eqnarray}\label{operat}
&&{\hat{b}^{\dagger}(m)}_{{l,j;}{[s]}}= \int d\vec{k}\frac{\delta\left(k-k_{m}\right)}{k}Y_{l,j}(\theta_{\vec{k}},\phi_{\vec{k}})\hat{a}^{\dagger}_{[s]}(\vec{k})\nonumber \\
&& \hat{a}^{\dagger}_{[s]}(\vec{k}) = \sum_{l,j}\int dm \frac{\delta(k_{m} -k)}{k_{m}} Y^{*}_{i,j}(\vec{k}) {\hat{b}^{\dagger}(m)}_{{l,j;}{[s]}}
\end{eqnarray}
where we introduced indices $[s]$ and $m$ denoting fixed value of polarization and value of LI mass and assume that $k_{m} = mc/\hbar$. One can easily check that operators ${\hat{b}^{\dagger}(m)}_{{l,j;}{[s]}}$ and ${\hat{b}(m)}_{{l,j;}{[s]}}$ obey bosonic commutation relations, where $m$ is assumed to be a continuous variable:
\begin{eqnarray}\label{syst_set}
&&\bigg[{\hat{b}(m)}_{{l,j;}{[s]}},{\hat{b}^{\dagger}(m)}_{{l,j;}{[s]}}\bigg] = \delta(m-m')\delta_{ll'}\delta{jj'}\nonumber \\
&&{\hat{b}^{\dagger}(m)}_{{l,j;}{[s]}}\ket{\rm{vac}} = \ket{m,lj}^{1-\rm{photon}}
\end{eqnarray}
The correspond states of ${\hat{b}^{\dagger}(m)}_{{l,j;}{[s]}}$ are normalized and form complete basis set:
\begin{eqnarray}
&&\braket{m,lj|{m'},{l'}{j'}} = \delta(m-m')\delta_{ll'}\delta{jj'} \nonumber \\
&&\sum_{l,j}\int dm\ket{m,lj}^{1-\rm{photon}}\bra{m,lj}^{1-\rm{photon}} = \hat{\mathds{1}}
\end{eqnarray}
We can generalize these states in case of multiphotons to standard Fock basis \cite{TG}.
Using \eqref{operat} and \eqref{operator} Hamiltonian takes a very simple form:
\begin{eqnarray}\label{Hamilt}
\hat{H}_{m} =\sum_{s}\int mc^2\left({\hat{b}^{\dagger}(m)}_{{l,j;}{[s]}}{\hat{b}(m)}_{{l,j;}{[s]}}\right)dk_{m},
\end{eqnarray}
where we add index $m$ to $\hat{H}$ stressing that we used non-standard modes.
However, for any $l,j$ and $m$, state $\ket{m,lj}$ is the eigenstate of Hamiltonian \eqref{Hamilt}:
$\hat{H}_{m}\ket{m,lj} = mc^2 \ket{m,lj}$. Constructed modes are a mathematical trick rather than a real physical description in some rest reference frame. We only rewrote the Hamiltonian using new sets of modes, but such operator does not possess Lorentz-invariance property nor even covariance. We can construct an exactly the same mode basis in any reference frame, but we must transform our multiphoton wave-packet state in accordance with Lorentz-transformations which affect both energy and momentum.
Nevertheless, presented approach gives simple classification of states with fixed value of LI mass. For instance, let us consider wave-packets $\ket{\Phi_{1}}$ and  $\ket{\Phi_{2}}$ that have the same value of LI mass, and even the same direction of mean propagation velocity in some laboratory reference frame, but  $\ket{\Phi_{1}^{\rm{l.f.}}} \neq e^{i\delta}\ket{\Phi^{\rm{l.f.}}_{2}}$, $\delta$ is an arbitrary phase. We can suppose that wave-packets have a common rest frame. Using \eqref{syst_set} we decompose our states:
\begin{eqnarray}\label{example}
&&\ket{\Phi_{1,2}}^{\rm{r.f.}} =\sum_{l,j} \int \braket{m,lj\Phi_{1,2}^{\rm{r.f.}}}\ket{m,lj}dm \nonumber \\
&&=\sum_{l,j}\beta_{{1,2}{l,j}}^{[m]}\ket{m,lj}; \nonumber \\
&&\beta_{{1,2}{l,j}}^{[m]} = \int {\phi_{1,2}\left(\vec{k}\right)}Y^{*}_{l,j}\left(\theta_{\vec{k}},\phi^{\rm{r.f.}}_{\vec{k}}\right)\frac{\delta(k-k_{m})}{k}d\vec{k},
\end{eqnarray}
where ${\phi_{1,2}^{\rm{r.f.}}\left(\vec{k}\right)}$ is an amplitude of wave-packets in the rest frame. Note that if the state has a fixed LI mass and $\vec{p} = 0$, that means that we can use orthogonality property, so integration over $k_{m}$  in \eqref{example} vanishes. We can write a scalar product of two states ($m$ is fixed value of LI mass): 
\begin{eqnarray}
\braket{\Phi^{\rm{r.f.}}_{1}\Phi^{\rm{r.f.}}_{2}} = \sum_{l,j}\beta_{{1}{l,j}}^{{*}[m]}\beta_{{2}{l,j}}^{[m]}.
\end{eqnarray}
Thus, spherical harmonic expansion coefficients $\beta_{{2}{l,j}}^{[m]}$ specify all possible states with given LI mass, as we proposed. One can decompose any function describing wave-packet state using this modes with well-defined LI mass according to \eqref{completeort} or expand any field operator. Theoretically, one can model a multiphoton wave-packet in a rest reference frame by using the basis set of states with well-defined LI mass, and then transform it back to the lab frame. 
Choosing value of the mean energy in lab reference frame to be $\braket{H}_{l.f.}$, we can calculate the relativistic gamma factor $\gamma = \braket{H}_{l.f.}/mc^2$ and make proper Lorentz transformation.
For a classical light pulse, the transition between reference frames is straightforward \cite{Landau}. Electric and magnetic field components are transformed as components of an electromagnetic field tensor under Lorentz transformation. 
In turn, the Lorentz transformation of a multiphoton wave-packet state is a little bit tricky. We recommend papers \cite{milburn} and \cite{han} for comprehensive details, where direct guidelines of how to calculate such transformation are given. Also, authors pointed out a problem which is related to the gauge freedom of the electromagnetic field. It is claimed that one can perform proper transformation by fixing the gauge and derive a special rule to correct transformation of polarization degree of freedom \cite{han}.
Specific examples of such transitions are beyond the scope of this paper and will be considered in further research.

\section{Discussion and Conclusion}\label{sec5}
Let us emphasize key results of the present work:
\begin{enumerate}
    \item We presented a physically reasonable prove that an arbitrary spatially localized wave-packet state propagates (in vaccum) with a mean velocity related to LI mass. More importantly, we emphasize the fact that mean propagation velocity is directly defined by mode-structures or one may say by geometry of an amplitude in $k$ - space.
    \item We conclude that the mean propagation velocity of an arbitrary multiphoton spatially localized wave-packet, which occupies finite spatial volume during propagation, is less than the speed of light constant in vacuum. This slowing-down effect can be observed in the experiment by the means of the intensity photodetection with a "bucket" detector, where the wave-packet state is registered as a whole.
    \item We provided an overview of the Lorentz invariant mass applicable to the multiphoton wave-packet states.
\end{enumerate}
It is worth to mention that it is challenging to construct LI mass operator based on \eqref{mass_general}, because operations $|{\rm{tr}}{\hat{\varrho} \hat{\vec{p}}}| \neq {\rm{tr}}{\hat{\varrho} |\hat{\vec{p}}|} $ are not generally commutative. Moreover fully relativistic treatment of localized wave-packet states is much more complicated. It gives rise of other nontrivial questions regarding photon position operators and position representations. For instance, we recommend works by M. Hawton \cite{H1,H2} for details.
From our analysis, it is clear that LI mass rather directly depends on the amplitude (modes) structure, but ties together corpuscular-wave nature of electromagnetic quanta. Also, we present a specific set of states with well defined LI mass where this corpuscular-wave nurture is clearly manifested. 

We believe that discussed effects can be effectively observed in experiment. For instance recently developed TimePix \cite{I_and_Andrei,Andrei1,Andrei2,Andrei3} camera allows one to detect time of arrival and measure spectral-temporal correlations using an array of approximately 65000 pixels, which is in fact a combination of independent detectors. In Section \ref{sec3} high orders of correlation functions are mentioned in context of possible enhancement of experimental setups and achieving high precision. Nowadays, such computationally complex task could be considered efficiently from the machine learning perspective. New methods arising from this field have already enabled a vast variety of applications in physics. Machine learning proved its efficiency in enhancement of traditional experimental techniques, for example \cite{kulik2020}. High order correlation analysis is very similar to image processing, which is one of the most developed areas of machine learning. In addition, TimePix camera mentioned above allows one to collect the required amount of training data, which is an essential part of any data driven approach. 

\section{Acknowledgement}
The work is supported by the Russian Foundation for Basic Research under grant no. 18-32-00906.  We thank Dr. Mikhail Fedorov and Savva Morozov for fruitful comments.

\section{Appendix}\label{appendix}
This part is aimed to provide the rigorous mathematical justification of results, described in Section \ref{sec3}.
We start our derivation by focusing on the intensity distribution at given point and time interval $t+dt$ for an arbitrary $n$-photon wave-packet state. We remind that intensity distribution is:
\begin{eqnarray}\label{intensity distr app}
p_{z}\left(\vec{r}_{\perp},t\right) = \frac{\Gamma(\vec{r},t,\vec{r},t)}{\int  \Gamma(\vec{r},t,\vec{r},t) d\vec{r}_{\perp}dt},
\end{eqnarray}
and
\begin{eqnarray}\label{velocity_int_app}
\braket{v} =|\braket{\frac{\partial \vec{r}}{\partial t}}| = |\int \frac{\partial \vec{r}}{\partial t}p_{z}\left(\vec{r}_{\perp},t\right)d\vec{r}_{{\perp}}dt|
\end{eqnarray}
The above expression \eqref{velocity_int_app} can be significantly simplified by integrating over $d\vec{r}_{\perp}dt$ as follows. Owing to $n$-photon state vector \eqref{WP_POL} does not depend on neither coordinates nor time and taking into account linearity of operators, one can swap integration over coordinates and the quantum mechanical averaging, namely $\int d\vec{r}dt \braket{\Psi A(\vec{r},t)\Psi} = \braket{\Psi (\int A(\vec{r},t) d\vec{r}dt)\Psi}$. At first, it is easy to notice that the denominator in \eqref{intensity distr app} is just the mean energy multiplied by coefficient $2\pi$:
\begin{eqnarray}\label{simpl}
&&\braket{\Psi\int \hat{\vec{E}}^{(-)}(\vec{r},t)\hat{\vec{E}}^{(+)}(\vec{r},t)d\vec{r}_{\perp}dt\ \Psi} = \bra{\Psi}\frac{\hbar c}{(2\pi)^2}\times\nonumber \\
&& \sum_{s,s'}\int d\vec{k}'d\vec{k}\bigg[ \vec{\varepsilon}^{*}_{{\vec{k}'}{s'}}\vec{\varepsilon}_{{\vec{k}}{s}}\sqrt{kk'}\hat{a}_{s'}^{\dagger}(\vec{k}')\hat{a}_{s}(\vec{k})\times \nonumber \\
&&e^{i((\vec{k} - \vec{k}')\vec{r}_{j}-(\omega_{k} - \omega_{k'})t_{j})}d\vec{r}_{{j}{\perp}}dt_{j}\bigg]\ket{\Psi} = \nonumber \\
&&(2\pi\hbar c)\braket{ \Psi\sum_{s}\int d\vec{k} k \hat{a}_{s}^{\dagger}(\vec{k})\hat{a}_{s}(\vec{k})\Psi} = 2\pi\braket{\hat{H}},
\end{eqnarray}
where we use well-known Dirac delta function integral representations: $\int e^{i(\vec{k} - \vec{k}')\vec{r}_{\perp}}d\vec{r}_{\perp} = (2\pi)^{2}\delta(\vec{k} - \vec{k'})$, $\int e^{-i(\omega_{k} - \omega_{k'})t}dt = 2\pi\delta(\omega_{k} - \omega_{k'})$, and relation $\omega_{k} = c\sqrt{k_{x}^2+k_{y}^2+k_{z}^2}$. The second simplification can be achieved in \eqref{velocity_int_app} by swapping again quantum averaging over state and integration over $d\vec{r}dt$ together with partial derivative $\frac{\partial \vec{r}}{\partial t}$. As a result we have to calculate: 
\begin{eqnarray}\label{derivatives}
&&\bra{\Psi}\int \frac{\partial \vec{r}}{\partial t}\hat{\vec{E}}^{(-)}(\vec{r},t)\hat{\vec{E}}^{(+)}(\vec{r},t) d\vec{r}dt\ket{\Psi} = \nonumber \\
&&\bra{\Psi}\int \frac{\partial}{\partial t}\left(\vec{r}\hat{\vec{E}}^{(-)}(\vec{r},t)\hat{\vec{E}}^{(+)}(\vec{r},t)d\vec{r}\right)dt - \nonumber \\
&& \int \vec{r}\left(\frac{\partial  \hat{\vec{E}}^{(-)}(\vec{r},t)\hat{\vec{E}}^{(+)}(\vec{r},t)}{\partial t}dt\right) d\vec{r}\ket{\Psi}
\end{eqnarray}
The expression at the bottom in \eqref{derivatives} vanishes. It is clear from exact view of field operators in \eqref{simpl}. Indeed, the differentiation over $t$ yields that integral vanishes $\int \dots -i(\omega_{k} - \omega_{k'})e^{-i(\omega_{k} - \omega_{k'})t}dt d\vec{k}d\vec{k'} = 0$ owing two subsequent integration over the time. Now, our main goal is to calculate integral $\int \frac{\partial}{\partial t}\left(\vec{r}\braket{\Psi\hat{\vec{E}}^{(-)}(\vec{r},t)\hat{\vec{E}}^{(+)}(\vec{r},t)\Psi}d\vec{r}\right)dt$.
Here we can use the following mathematical trick. Let us specify direction of $\vec{r} = |{r}_{\perp}|\vec{e}_{\perp} + z\vec{e}_{z} $ in fixed coordinate system $\{\vec{e}_{\perp}, \vec{e}_{z}\}$. We substitute $\vec{r}$ in \eqref{derivatives} and use explicit expressions of the field operator \eqref{operator} in \eqref{derivatives} on the next step. Now, we obtained two terms multiplied by basis vectors $\vec{e}_{\perp}$ and $\vec{e}_{z}$:
\begin{eqnarray}\label{z-and-r-int}
&&\sum_{s,s'}\int \braket{\Psi\hat{a}_{s'}(\vec{k'})^{\dagger}\hat{a}_{s}(\vec{k})\Psi}\dots\int \left(\vec{r}_{\perp}e^{i(\vec{k}_{\perp} - \vec{k'}_{\perp})\vec{r}_{\perp}}d\vec{r}_{\perp}\right)\times\nonumber \\
&&e^{-i(\omega_{k} - \omega{k'})t}e^{i(k_{z}-k'_{z})z}d\vec{k}_{\perp}d\vec{k'}_{\perp}dk_{z}dk'_{z}+ \nonumber\\
&&\vec{e}_{z}\sum_{s,s'}\int \braket{\Psi\hat{a}_{s'}(\vec{k'})^{\dagger}\hat{a}_{s}(\vec{k})\Psi}\dots \left(ze^{i(k_{z} -k'_{z})z}\right) \times \nonumber\\
&&\delta(\vec{k}_{\perp} - \vec{k'}_{\perp})e^{-i(\omega_{k} - \omega{k'})t}d\vec{k}_{\perp}d\vec{k'}_{\perp}dk_{z}dk'_{z}
\end{eqnarray}
To avoid cumbersome expressions further we will write down only the most important terms. 
Let us modify expression above \eqref{z-and-r-int} by rewriting the two  following terms:
\begin{eqnarray}\label{mod}
&&1: ze^{i(k_{z} -{k'}_{z})z}e^{-i(\omega_{k} - \omega{k'})t} = \nonumber \\
&&\frac{\partial}{i\partial(k_{z} -{k'}_{z})}\bigg[e^{i(k_{z} -{k'}_{z})z}e^{-i(\omega_{k} - \omega_{k'})t}\bigg] - \nonumber \\
&&e^{i(k_{z} -{k'}_{z})z}\bigg[\frac{\partial}{i\partial(k_{z} -{k'}_{z})}e^{-i(\sqrt{k_{\perp}^2+k_{z}^2} - \sqrt{{k'}_{\perp}^2+{k'}_{z}^2})ct}\bigg]\nonumber \\ 
&&2: \int \vec{r}_{\perp}e^{i(\vec{k}_{\perp} - \vec{k'}_{\perp})\vec{r}_{\perp}}d\vec{r}_{\perp} =
\frac{\partial \delta(\vec{k}_{\perp} - \vec{k'}_{\perp})} {i\partial(\vec{k}_{\perp} -\vec{k'}_{\perp})}. 
\end{eqnarray}
We already noticed after eq.\eqref{derivatives} that the subsequent differentiation and integration over time gives zero value, if terms like
$(\omega_{k}-\omega_{k'})e^{-i(\omega_{k} - \omega_{k'} + \dots)t}$ are presented in the integrals. Moreover, as far as coordinates $z$ and $r_{\perp}$ depends on time linearly and does not lead to confusion by adding non-linear dependence on time. 
Thus, it is clear that first term in expression 1: \eqref{mod} gives non-zero value after it substitution in \eqref{z-and-r-int} and then subsequent time differentiation and integration.
Here we implemented the well-known property of a delta-function $\int f(y)\frac{\partial}{\partial y}\delta(y-y_{0})dy = -\frac{\partial f(y_{0})}{\partial y_{0}}$. Indeed, utilizing this this property, we get the following term: $ct\bigg[\frac{\partial\left(\sqrt{k_{\perp}^2+k_{z}^2} - \sqrt{{k'}_{\perp}^2+{k'}_{z}^2}\right)}{\partial \left(k_{z} - {k'}_{z}\right)}\bigg]e^{-i\left(\sqrt{k_{\perp}^2+k_{z}^2} - \sqrt{{k'}_{\perp}^2+{k'}_{z}^2}\right)ct}$. The presence of this terms in integrals leads to nonzero results after time differentiation-integration over time $t$ and integration over variables $\vec{k},\vec{k'}$. Somewhat similar situation occurs regarding term 2 in \eqref{mod}. Here, term like $ ct\bigg[\frac{\partial\left(\sqrt{\vec{k}_{\perp}^2 + {k}_{z}^2} - \sqrt{\vec{k'}_{\perp}^2 + {k'}_{z}^2}\right)}{\partial \left(\vec{k}_{\perp} - \vec{k'}_{\perp}\right)}\bigg]e^{-i(\omega_{k} - \omega_{k'} + \dots)t}$ occurs in \eqref{z-and-r-int}.
To perform differentiation let us make variable changes:
\begin{eqnarray}\label{trick1}
&& \vec{k}_{\perp} - \vec{k'}_{\perp} = \vec{\eta}_{-}; \ \vec{k}_{\perp} + \vec{k'}_{\perp} = \vec{\eta}_{+} \nonumber\\
&&k_{\perp}^2 = (1/4)\left(\eta_{-}^2+\eta_{+}^2 + 2\eta_{-}\eta_{+}\cos(\alpha)\right);   \nonumber\\
&&{k'}_{\perp}^2 = (1/4)\left(\eta_{-}^2+\eta_{+}^2 - 2\eta_{-}\eta_{+}\cos(\alpha)\right);
\end{eqnarray}
In turn, for $k_{z} - k'_{z}$ we have: 
\begin{eqnarray}\label{trick2}
&&k_{z} - k'_{z} = \xi_{-} \nonumber\\
&& k_{z} + k'_{z} = \xi_{+}.
\end{eqnarray}
For the term $ \frac{\partial\left(\sqrt{\vec{k}_{\perp}^2 + {k}_{z}^2} - \sqrt{\vec{k'}_{\perp}^2 + {k'}_{z}^2}\right)}{\partial \left(\vec{k}_{\perp} - \vec{k'}_{\perp}\right)}$ after variable change we get:
\begin{eqnarray}\label{almost_done}
&&\frac{\partial\left(\sqrt{k_{\perp}^2(\vec{\eta_{-}},\vec{\eta_{+}})+k_{z}^2} - \sqrt{{k'}_{\perp}^2(\vec{\eta_{-}},\vec{\eta_{+}})+{k'}_{z}^2}\right)}{\partial \vec{\eta}_{(-)}} \nonumber \\
 &&= \frac{\frac{1}{2}\vec{\eta_{+}}}{\sqrt{(1/4)\eta_{+}^2+k_{z}^2}}|_{\vec{\eta}_{-}\equiv 0} = \frac{\vec{k}_{\perp}}{k}
\end{eqnarray}
Identical for the term differentiated over $k_{z} - {k'}_{z}$ yields:
\begin{eqnarray}\label{almost_done2}
\frac{\partial\left(\sqrt{k_{\perp}^2+k_{z}^2} - \sqrt{{k'}_{\perp}^2+{k'}_{z}^2}\right)}{\partial \left(k_{z} - {k'}_{z}\right)} = \frac{k_{z}}{k}
\end{eqnarray}
Note that we took derivatives over $\eta_{-}$ and $\xi_{-}$ at points where $\eta_{-}$ and $\xi_{-}$ are zero, consequently it means that $\vec{k}_{\perp} = \vec{k'}_{\perp}$ and $k_{z} = {k'}_{z}$. Finally, substituting \eqref{mod} and \eqref{z-and-r-int} in \eqref{derivatives}, excluding all zero-valued terms, using the explicit form of $n$-photon state $\ket{\Psi}$ in \eqref{WP_POL} and performing long but straightforward algebraic manipulations we finally get: 
\begin{eqnarray}
&&\int \frac{\partial}{\partial t}\left(\vec{r}\braket{\Psi\hat{\vec{E}}^{(-)}(\vec{r},t)\hat{\vec{E}}^{(+)}(\vec{r},t)\Psi}d\vec{r}\right)dt = \nonumber\\
&&c\int (k_{\perp}\vec{e}_{\perp}+k_{z}\vec{e}_{z}) \sum_{s} |\psi_{s}\left(\vec{k}\right)|^2 d\vec{k} = c\braket{\hat{\vec{p}}} 
\end{eqnarray}
Therefore, combining all results above we prove that 
\begin{eqnarray}
\braket{v} =|\braket{\frac{\partial \vec{r}}{\partial t}}| = c\frac{|\braket{\hat{\vec{p}}}|}{\braket{H}}
\end{eqnarray}
%\newpage

\clearpage

\end{document}